\documentclass[12pt]{article}
\usepackage{a4wide,feynmf,amsmath,amssymb,graphicx,psfrag}

\allowdisplaybreaks[1]

\begin{document}\setlength{\unitlength}{1mm}


\newcommand{\wt}{\widetilde}

\newcommand{\bra}[1]{\langle #1|}
\newcommand{\ket}[1]{|#1\rangle}
\newcommand{\bracket}[2]{\bra{#1}#2\rangle}
\newcommand{\erw}[1]{\langle #1\rangle}
\newcommand{\kom}[2]{\left[#1,#2\right]}
\newcommand{\akom}[2]{\left\{#1,#2\right\}}
\newcommand{\norm}[1]{\parallel #1\parallel}

\newcommand{\vsp}{\vspace{1cm}}
\newcommand{\svsp}{\vspace{0.5cm}}
\newcommand{\hsp}{\hspace{1cm}}
\newcommand{\shsp}{\hspace{0.5cm}}
\newcommand{\ehsp}{\hspace{0.25cm}}

\newcommand{\E}[1]{\cdot 10^{#1}}

\newcommand{\spur}[1]{\backslash\, #1 \,\backslash}

\newcommand{\ddk}{d^{D}\!k}
\newcommand{\iddk}{\int\!\!\ddk}
\newcommand{\mupi}{\frac{\mu^{4-D}}{(2\pi)^D}}
\newcommand{\dvk}{d^4\!k}
\newcommand{\idvk}{\int\!\!\dvk}
\newcommand{\idvkzpi}{\int\!\!\frac{\dvk}{(2\pi)^4}}

\newcommand{\dmuu}{\partial_\mu}
\newcommand{\dmuo}{\partial^\mu}
\newcommand{\dnuu}{\partial_\nu}
\newcommand{\dnuo}{\partial^\nu}

\newcommand{\ve}{\varepsilon}

\newcommand{\polv}[3]{\varepsilon_{#1}^{#2}(#3)}
\newcommand{\PolTensor}
	{\varepsilon_{a,\lambda}^{\mu}(p_1)\varepsilon_{b,\lambda'}^{\nu}(p_2)}


\newcommand{\adot}{ {\dot{\alpha}} }
\newcommand{\bdot}{ {\dot{\beta}} }
\newcommand{\gdot}{ {\dot{\gamma}} }
\newcommand{\deldot}{ {\dot{\delta}} }
\newcommand{\thau}{ \theta^\alpha } 
\newcommand{\thbau}{ \bar{\theta}^\adot }
\newcommand{\thad}{ \theta_\alpha }
\newcommand{\thbad}{ \bar{\theta}_\adot }

\newcommand{\sigmuaa}{(\sigma^\mu)_{\alpha\adot}}

\newcommand{\TT}{\theta\theta}
\newcommand{\TBTB}{\bar{\theta}\bar{\theta}}
\newcommand{\HK}{\text{h.k.}}

\newcommand{\gev}{\text{GeV}}
\newcommand{\tev}{\text{TeV}}
\newcommand{\fb}{\text{fb}}
\newcommand{\pb}{\text{pb}}


\newcommand{\MM}{{\cal M}}

\newcommand{\Sq}{   {\tilde{\text{q}}} }
\newcommand{\Stop}{ {\tilde{\text{t}}} }
\newcommand{\Sbot}{ {\tilde{\text{b}}} }

\newcommand{\gghh}{$g g \to H^+ H^-$}

\newcommand{\mhpm}{ m_{H^{\pm}} }
\newcommand{\msqu}{M_{\text{squark}}}
\newcommand{\MSQ}{ M_{\tilde{Q}} }
\newcommand{\MSU}{ M_{\tilde{U}} }
\newcommand{\MSD}{ M_{\tilde{D}} }
\newcommand{\sqrts}{ \sqrt{s} }

\newcommand{\gshpmtb}{g_s [ H^{\pm}, t, b ]}
\newcommand{\gphpmtb}{g_p [ H^{\pm}, t, b ]} 
\newcommand{\ghiqq}{g [H^0_i, q, q ]}
\newcommand{\ghihphm}{g [H^0_i, H^{+}, H^{-}]}
\newcommand{\gsqlsqlhi}{g [H^0_i, \Sq_l, \Sq_l]}
\newcommand{\gsqlsqlhphm}{g [\Sq_l, \Sq_l, H^{+}, H^{-}]}
\newcommand{\ghpasbottomastope}[2]
           {g [ H^{+}\; aus,\Stop_{#2}\; ein,\Sbot_{#1}\; aus ] }
\newcommand{\ghmastopasbottome}[2]
           {g [ H^{-}\; aus,\Sbot_{#2}\; ein,\Stop_{#1}\; aus ] }
\newcommand{\ghpmstopsbottom}[2]{ g [ H^{\pm}, \Stop_{#1}, \Sbot_{#2} ] }


\begin{titlepage}
\begin{flushright}
KA--TP--11--1999\\
hep-ph/9908529\\
August 1999
\end{flushright}
\vspace{2cm}
\begin{center}
{\Large Pair production of charged MSSM Higgs bosons by gluon fusion} \\[2.5cm]
{\large O.~Brein \footnote{E-mail: obr@itp.uni-karlsruhe.de},
W.~Hollik \footnote{E-mail: Wolfgang.Hollik@physik.uni-karlsruhe.de}} \\[0.8cm]
{\normalsize\em 
	Institut f\"ur Theoretische Physik, Universit\"at Karlsruhe,\\
        D-76128 Karlsruhe, Germany}
\end{center}
\vspace{2.5cm}
\begin{abstract}
The production of pairs of charged Higgs bosons as predicted by the 
minimal supersymmetric standard model (MSSM) via the gluon fusion
mechanism is investigated. 
The amplitudes at the leading one-loop order for the parton process
$gg \to H^+H^-$
are calculated with the complete set of MSSM particles. 
Numerical results are presented for the cross section of the inclusive
hadron process $p p \to gg \to H^+ H^- + X$ at the LHC. 
\end{abstract}

\end{titlepage}


\section{Introduction}

The search for Higgs bosons and the study of the mechanism for
the electroweak symmetry breaking is a basic task for future
high energy colliders.
Charged Higgs particles are part of an extended scalar sector
with at least two isodoublets of Higgs fields. Such a minimal
scenario with two Higgs doublets is realized in the 
minimal supersymmetric standard model (MSSM). A possible detection
of charged Higgs particles would therefore be a clear signal for
the presence of non-standard physics, with a strong hint towards
supersymmetry.
For an accurate identification a precise determination of the 
production and decay properties of the non-standard particles is 
a crucial requirement. 
At hadron colliders, $H^+H^-$ pairs can be produced
by two basic mechanisms: annihilation of quark-antiquark pairs
and gluon-gluon fusion. The gluon fusion mechanism is 
particularly interesting since, as a loop-induced higher-order process,
it depends on all virtual particles with couplings to 
gluons as well as to Higgs bosons. In the MSSM, at the
one-loop level,  the process is mediated through quarks and
squarks (predominantly from the third generation), involving also neutral
Higgs bosons and the Higgs self-interaction.    
Although the cross section for $H^+H^-$ production 
by $q\bar{q}$ annihilation \cite{EHLQ,willenbrock} is larger, 
it is less sensitive
to the details of the underlying model than the gluon fusion mechanism.
In particular, through the virtual presence of scalar quarks,
the MSSM cross section in general is  different from the cross section
for a non-supersymmetric two-doublet-model even with identical Yukawa
couplings. 
At the LHC, 
the supression of the  cross section for $gg\to H^+H^-$
by a factor $\alpha_S^2$ compared to
$q\bar{q}\to H^+H^-$  
is partly compensated by the high gluon luminosity.

Gluon fusion into charged Higgs particles
has been studied previously in \cite{KPSZ} 
based on the loop contributions from
top and bottom quarks. Scalar quark contributions are 
neglected, which corresponds to the decoupling scenario of
sufficiently heavy squarks. 
In this article we extend the calculation by including also
a non-decoupling 
scalar quark sector in the loop diagrams; we give 
analytical results and 
a detailed discussion for the size of the various contributions.
In a quite recent paper \cite{kniehl} 
the squark loops to the gluon fusion process have also been derived.
We find agreement with our results.

The paper is organized as follows:
After a short description of the MSSM entries for the calculation of the 
$H^+H^-$ cross section, we derive the parton cross section
in Section 3. Section 4 contains the hadronic cross section and 
a numerical analysis. The appendix collects the analytical
expressions, the list of coupling constants, and the Feynman graphs
for the gluon fusion process.

\section{MSSM entries}

The cross section for $H^+H^-$ production by the 
lowest order Drell-Yan mechanism depends 
only on the mass $\mhpm$ and the gauge couplings of the charged Higgs
boson. $H^+H^-$ pair production via gluon fusion, as a higher-order
process merging the strong and electroweak interactions, depends on 
the detailed structure of the Higgs and the quark/squark sector 
of the MSSM.

\subsection{Higgs sector}
Besides the charged Higgs particles $H^{\pm}$, the MSSM Higgs sector
contains the two neutral CP-even scalars
\footnote{The notation $H^0_1,\,H^0_2$
is equivalent to $H^0,\, h^0$
in the conventions of \cite{HHG}.} 
$H^0_1,H^0_2$
and the neutral CP-odd `pseudoscalar' $A^0$
as physical mass eigenstates.
The free parameters of the Higgs sector are frequently chosen to 
be the mass $m_A$ of the $A^0$ boson
and the ratio $\tan\beta = v_2/v_1$ of the vacuum 
expectation values of the two Higgs doublets.  
For the present analysis it is  more convenient to choose the 
mass $\mhpm$ of 
the charged particles instead of $m_A$ 
as the free mass parameter. The relation 
between $m_{H^\pm}$ and $m_A$ in lowest order is given by
\begin{align}
\label{chargedmass}
m_{H^\pm}^2 & = m_W^2 + m_A^2 
\end{align}
with the $W$ mass $m_W$.
The CP-even Higgs masses, at the tree level, 
are given by:
 \begin{align}\label{mssm-higgsmassen}
m^2_{H^0_1,H^0_2} & = \frac{1}{2} 
\bigg[ 
m^2_{A} + m^2_Z \pm \sqrt{(m^2_{A} + m^2_Z)^2 
        - 4 m^2_Z m^2_{A}\cos^2 2\beta}
\bigg] 
\ehsp.
\end{align}
The angle $\alpha$ of the rotation  diagonalizing the 
neutral CP-even mass matrix, an entry in several coupling constants,
fulfills the tree level relation
\begin{align}
\label{alpha}
\tan 2\alpha & = \frac{m_H^2 + m_h^2}{m_A^2 - m_Z^2 } \tan 2 \beta 
\ehsp.
\end{align}
Both relations (\ref{mssm-higgsmassen}) and (\ref{alpha})
are sizeably modified by quantum corrections, predominantly proportional
to $m_t^4$. They are taken into account in the approximation given
in \cite{ERZ}. 
The relation (\ref{chargedmass})
between $m_A$ and $\mhpm$ is also modified
by radiative corrections \cite{Diaz}; the effects, however, are essentially
smaller and are neglected in our analysis.

\subsection{Squark sector}


The main entries from the scalar quark sector are from the
third generation of top ($\Stop_L,\Stop_R$) and
bottom ($\Sbot_L,\Sbot_R$) squarks.
For simplicity we assume a flavor-diagonal squark sector.
For the $\Stop$ and $\Sbot$ case
the mass matrices have the form

\begin{equation}\label{M-stop}
M_{\Stop}^2 =
\left(
\begin{array}[2]{cc}
M_{\tilde{Q}}^2+m_t^2+m_Z^2(\frac{1}{2}-e_t s_w^2)\cos2\beta &
m_t(A_t - \mu \cot\beta) \\
m_t(A_t - \mu \cot\beta) &
M_{\tilde{U}}^2+m_t^2+m_Z^2 e_t s_w^2\cos2\beta\\
\end{array}
\right)         \ehsp,
\end{equation}

\begin{equation}\label{M-sbottom}
M_{\Sbot}^2 =
\left(
\begin{array}[2]{cc}
M_{\tilde{Q}}^2+m_b^2+m_Z^2(-\frac{1}{2}-e_b s_w^2)\cos2\beta &
m_b(A_b - \mu \tan\beta) \\
m_b(A_b - \mu \tan\beta) &
M_{\tilde{D}}^2+m_b^2+m_Z^2 e_b s_w^2\cos2\beta\\
\end{array}
\right)         \ehsp,
\end{equation}
and analogously for the other generations.
The symbols $e_t$ and $e_b$ denote the electric charge of top- and
bottom-quarks;
$\mu$ is the supersymmetric Higgs mass parameter
\footnote{Our sign convention for $\mu$ is opposite 
          to that of \cite{HHG}.},
$M_{\tilde{Q}}$ the soft-breaking mass parameter for the 
squark isodoublet $(\Stop_L, \Sbot_L)$, and
$M_{\tilde{U}}$ and $M_{\tilde{D}}$ are the soft-breaking 
mass parameters for the isosinglets $\Stop_R$ and $\Sbot_R$.
They can be different for each generation, but for simplicity
we will assume equal values for all generations in our numerical analysis.
$A_t$ and $A_b$ are the parameters of the soft-breaking scalar three-point 
interactions of top- and bottom-squarks with the Higgs fields.

We restrict our analysis to real parameters. Therefore the
mass matrices for $\Stop$ and $\Sbot$
are real and can be diagonalized by means of rotations with angles
$\theta_{\Stop}$ and $\theta_{\Sbot}$.
The mass eigenstates 
$\Stop_1$ and $\Stop_2$ are obtained as
\begin{equation*}
\left(
\begin{array}[1]{c}
\Stop_1\\
\Stop_2
\end{array}
\right)
=
\left(
\begin{array}[2]{cc}
\cos\theta_{\Stop} & -\sin\theta_{\Stop}\\
\sin\theta_{\Stop} &  \cos\theta_{\Stop}
\end{array}
\right)
\left(
\begin{array}[1]{c}
\Stop_L\\
\Stop_R
\end{array}
\right) \, ,
\end{equation*}
and analogously for $\Sbot_1$ and $\Sbot_2$. 

\bigskip
For the pair production of charged Higgs particles via gluon fusion
all couplings of charged and neutral CP-even Higgs particles to squarks
appear at leading order. The couplings for mass eigenstates are 
listed in Appendix \ref{couplings}.

\section{The parton process $gg \to H^- H^+$}

The production of charged Higgs pairs by gluon fusion in the MSSM 
is a loop-induced process, at leading (1-loop) order 
mediated through quark and squark loops.
The Feynman graphs are depicted in Appendix \ref{feynman-graphs}. 

\bigskip
In our kinematical conventions,
the momenta of the initial and final state particles are all chosen as
incoming:
\[
 g(p_1,a,\lambda) + g(p_2,b,\lambda')  
  \rightarrow H^-(-p_3) + H^+(-p_4) \, .
\]
Besides by their momenta, the initial state gluons are 
characterized by their colour indices $a,b$ and their helicities
$\lambda,\lambda'$.
We make use of the parton kinematical invariants
\begin{align*}
\hat{s}  = (p_1+p_2)^2\, , \quad 
\hat{t}  = (p_1+p_3)^2\, , \quad
\hat{u}  = (p_1+p_4)^2
\end{align*}
obeying the relation
\begin{align*}
\hat{s} + \hat{t} + \hat{u} = 2 \mhpm^2
\ehsp.
\end{align*}
The spin- and colour-averaged cross section for the parton process
\begin{align}
\frac{d \sigma}{ d \hat{t}} & = \frac{1}{16\pi \hat{s}^2} \:
\frac{1}{4} \sum_{\lambda,\lambda' = \pm 1}^{}
\frac{1}{64} \sum_{a,b = 1}^{8}
\big| \MM_{\lambda\lambda'}^{ab} \big|^2
\end{align}
contains the helicity amplitudes
\begin{align}
\MM_{\lambda\lambda'}^{ab} & =
\polv{a\lambda}{\mu}{p_1} \polv{b\lambda'}{\nu}{p_2} \,
\widetilde{\MM}_{\mu\nu} \cdot
(4\pi \alpha_S)\,
\text{Tr} \bigg\{ \frac{\lambda^a}{2} \frac{\lambda^b}{2} \bigg\}
\ehsp,
\end{align}
where a pure QCD factor is written separately.
The remaining tensor $\widetilde{\MM}_{\mu\nu}$
can be decomposed  
with respect to a complete set of 
orthogonal tensors $A_i^{\mu\nu}$ \cite{GB,PSZ} according to
\begin{align}
\widetilde{\MM}^{\mu\nu} = 
\sum_{i=1}^{4} F_i(\hat{s},\hat{t}) A_i^{\mu\nu} (\hat{s},\hat{t})
\ehsp.
\end{align}
The explicit formulae for the basic tensors
are listed in Appendix \ref{tensorbasis}.
The tensor  $\widetilde{\MM}_{\mu\nu}$ is transverse:
\begin{align}
p_1^\mu \widetilde{\MM}_{\mu\nu} = 0\, , \quad
& p_2^\nu \widetilde{\MM}_{\mu\nu}  =0 \ehsp,
\end{align}
and the $A_i^{\mu\nu}$ 
are chosen to be individually transverse as well.
This transversality guarantees 
that the spin-averaged cross section is proportional to
$\widetilde{\MM}^{\mu\nu}\widetilde{\MM}^{\star}_{\mu\nu}$ and,
together
with the orthogonality of the basis tensors, that the differential 
cross section for the parton process can be expressed in a simple
way in terms of the form factors:
\begin{align}
\frac{d \sigma}{ d \hat{t}} & = \frac{\pi \alpha_S^2}{64\,\hat{s}^2}
\sum_{i=1}^{4} \big| F_i(\hat{s},\hat{t}) \big|^2
\ehsp.
\end{align}
The form factors can be divided into contributions from loop diagrams
with virtual squarks ($\Sq$) and quarks ($\text{q}$) 
arising in  2-, 3- and 4-point integrals
(denoted by the symbols $\circ$ for 2, $\triangle$ for 3,
and $\square$ for 4 internal lines):
\begin{align}
\nonumber
F_1 & = F^{\Sq,\circ} + F^{\Sq,\triangle} 
        + F^{\Sq,\square} 
        + F^{\text{q},\triangle}  + F^{\text{q},\square} \\
F_2 & = G^{\Sq,\circ} + G^{\Sq,\triangle} 
        + G^{\Sq,\square} 
        + G^{\text{q},\square} \\
\nonumber
F_3 & = 0\\
\nonumber
F_4 & = H^{\text{q},\square} \ehsp.
\end{align}
The tensor $A_3^{\mu\nu}$ projects on a CP-odd gluon state 
with a total spin component $ =0$ along the collision
axis in the CMS. Since the final state
is CP-even for this spin configuration, $F_3$ vanishes identically.
Analytic formulae for the form factors in terms of 
scalar 3- and 4-point-functions are presented in Appendix
\ref{formfactors}. The analytic formulae for the form factors of the 
quark loop graphs are in agreement with \cite{KPSZ}. For 
a complete and coherent documentation  
they are also included in Appendix \ref{formfactors}.

\bigskip
The integrated parton cross section
\begin{align}
\sigma_{gg \to H^+ H^-}(\hat{s},\alpha_S(\mu_R)) & =
\int_{ \hat{t}_{\text{min}}(\hat{s}) }^{ \hat{t}_{\text{max}}(\hat{s}) } 
\!\!\! d\hat{t}
\: \frac{d \sigma}{ d \hat{t}}
\end{align}
is evaluated by numerical integration over the 
kinematically allowed $\hat{t}$-range 
for a given CMS-energy $\sqrt{\hat{s}}$ of the gluon-gluon
system.
The renormalization
scale $\mu_R$ in the strong coupling
constant is chosen as $\sqrt{\hat{s}}$, the hard energy scale of the process.

\begin{figure}[hbt]
\begin{center}
  \psfrag{XL}[c]{$\sqrt{\hat{s}} \; [\gev]$}
  \psfrag{YL}[c]{$\sigma [\fb]$}
  \psfrag{MSQ100}{\scriptsize $M_{\text{Squark}} = 100\,\gev$}
  \psfrag{MSQ200}{\scriptsize $M_{\text{Squark}} = 200\,\gev$}
  \psfrag{MSQ400}{\scriptsize $M_{\text{Squark}} = 400\,\gev$}
  \psfrag{MSQ1000}{\scriptsize $M_{\text{Squark}} = 1000\,\gev$}
  \psfrag{SIGMAGGHPHMX}[l]{$\sigma ( gg \to H^+ H^- + X),\,\tan\beta = 1.5$}
  \includegraphics*{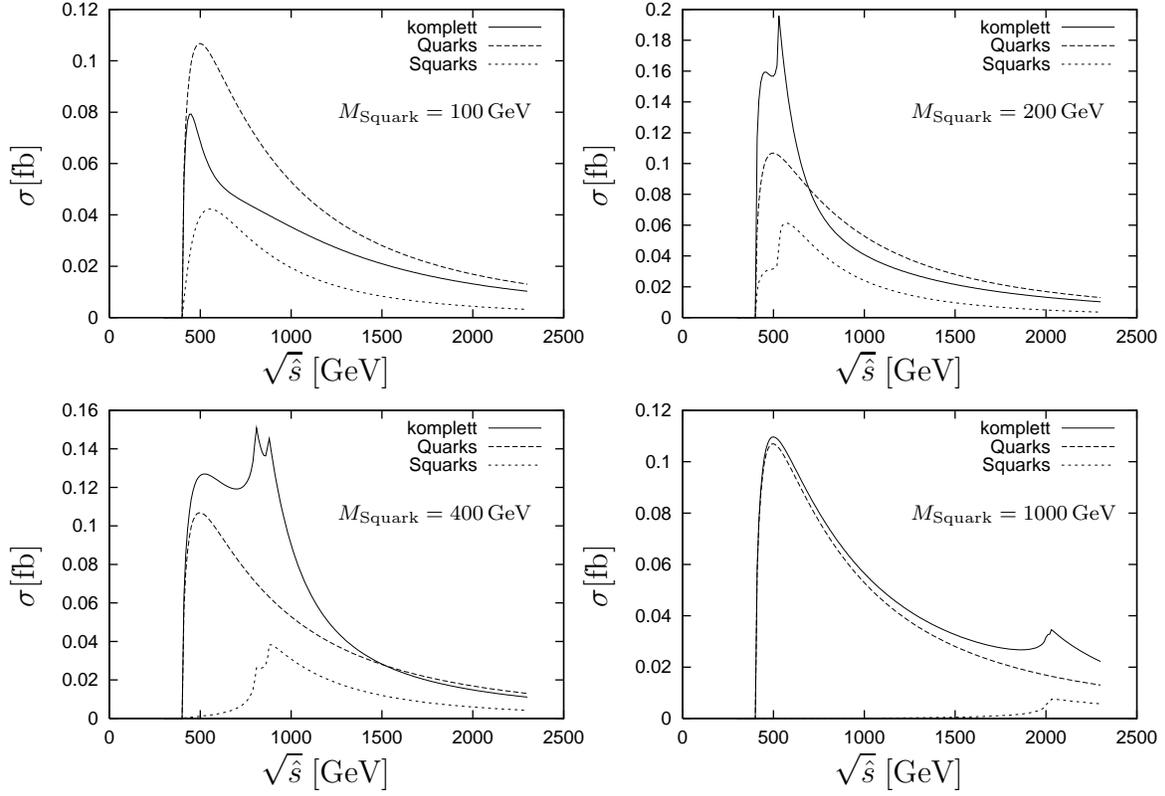}
    \caption{\label{pwq.s} \em 
	Parton cross section for the production of charged Higgs pairs
	($\mhpm = 200\,\gev$,$\tan\beta = 1.5$)
	via gluon fusion for rising values of the common squark mass
	scale ($100/200/400/1000\,\gev$). The cross section is evaluated
	with all Feynman graphs (solid lines), only quark loop graphs
	(long dashed) and only squark loop graphs (short dashed).
            }
  \end{center}
\end{figure}

\bigskip
In Figure \ref{pwq.s} the parton cross section is displayed for 
unmixed squarks ($A=\mu=0$) for 
four characteristic values of the squark mass scale which we
assume to be common for the various entries in the mass matrices:
$M_{\tilde{Q}} = M_{\tilde{U}} = M_{\tilde{D}}=\,: \msqu$. 
The parameters of the Higgs sector are $\mhpm = 200\,\gev$ and 
$\tan\beta = 1.5$. The Figure contains a sequence of graphs with
rising values of the squark mass scale, beginning with squark
masses below $\mhpm$, 
then slightly above $\mhpm$, followed by 
squark masses around $2 \mhpm$, and finally 
 at $1000\,\gev$.
The decoupling property of heavy squarks is clearly visible,
but it also becomes clear that squarks of intermediate 
mass scales can give 
rise to a sizeable enhancement of the cross section.
Due to significant interferences, the 
cross sections including both
quark and squark loops are not simply the sum of the individual
contributions.
The spikes in the plots correspond to
the squark pair thresholds of the loop diagrams.

\section{The hadron process $pp \to H^- H^+ + X$}

\subsection{Total hadronic cross section}\label{total-hadron-cs}

The hadronic cross section for $H^\pm$ pair 
production from gluons in proton-proton 
collisions at a total hadronic CMS energy $\sqrt{S}$
can be written as a convolution \cite{pQCD-lect}
\begin{align}
\label{hadronx}
\sigma \, \equiv \,
\sigma( p p  \to H^+ H^- + X) & = 
\int_{\tau_0}^{1} d\tau \frac{ d{\cal L}^{pp}_{gg} }{ d\tau }
\;\sigma_{gg\to H^+ H^-}(\tau S,\alpha_S(\mu_R))
\end{align}
with the gluon luminosity
\begin{align}
\frac{ d{\cal L}^{pp}_{gg} }{ d\tau } & = 
        \int_{\tau}^{1} \frac{dx}{x} 
        f_{g/p} (x,\mu_F) f_{g/p} (\frac{\tau}{x},\mu_F)\, ,
\end{align}
where $f_{g/p} (x,\mu_F)$ denotes the density of gluons
in the proton carrying
a fraction $x$ of the proton momentum
at the scale $\mu_F$.
The numerical evaluation has been carried out with the MRS(G)
gluon distribution functions \cite{MRSG} and with the renormalization
and factorization scale $\mu_R,\mu_F$ chosen as equal.

\begin{figure}[htb]
\begin{center}
  \psfrag{XL}[l]{$\tan\beta$}
  \psfrag{YL}[b]{$\sigma [\fb]$}
  \psfrag{SIGMAPPGGHPHMX}[l]{$\sigma ( pp \to H^+ H^- + X)$}
  \resizebox*{1.0\width}{1.0\height}{\includegraphics*{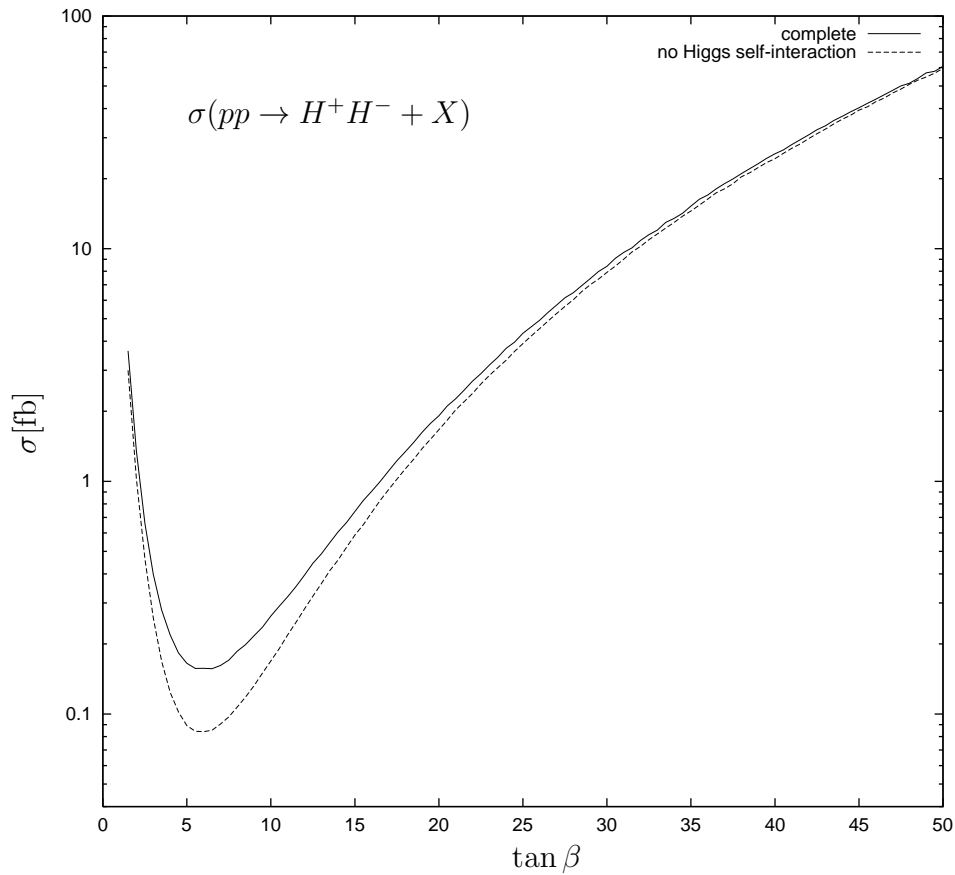}}
    \caption{\label{tb.mhpm200.msq200.all-no3H} 
	\em Variation of the hadronic cross section with $\tan\beta$
	calculated with (solid line) and without (dashed line) the 
	trilinear Higgs couplings. The mass of the charged 
	Higgs is $\mhpm = 200\,\gev$ and the squark mass scale 
	is $\msqu = 200\,\gev$.
	}
  \end{center}
\end{figure}

\bigskip
Table \ref{wq-zahlen} collects a sample of 
numerical results for the hadronic cross section for typical sets
of MSSM parameters.
Quark masses are $m_t = 175\,\gev$, $m_b = 5\,\gev$.
We always assume a common squark mass scale 
$\msqu (= M_{\tilde{Q}} = M_{\tilde{U}} = M_{\tilde{D}})$, 
which in table \ref{wq-zahlen} has been set 
to 200 $\gev$ (where the squark diagrams give a big contribution)
 and to infinity describing the case of decoupling squarks. 
An unmixed squark mass scenario is assumed ($A=\mu=0$).
$\tan\beta = 6$ is chosen as a reference point for the  
minimal cross section (see also Figure \ref{tb.mhpm200.msq200.all-no3H}),
and $\tan\beta = 50$
as a representative value for the large $\tan\beta$ scenario.

\begin{table}[hbt]
\begin{center}
\begin{tabular}[7]{|c|c|c|c|c|c|c|}
\hline
$\tan\beta$ & $\mhpm$ [$\gev$] & $\msqu$ [$\gev$] 
        & \multicolumn{2}{|c|}{$\sqrt{S}$ = 14 $\tev$} 
        & \multicolumn{2}{|c|}{$\sqrt{S}$ =  2 $\tev$} \\ \cline{4-7}
&&& $\sigma$ [$\fb$] & events & $\sigma$ [$\fb$] & events \\
\hline
\hline
50 & 100 & 200      & 140.7 & $\approx$ 14000 & 0.48      & $\approx$ 1\\ 
\cline{3-7}
   &     & $\infty$ & 79.7  & $\approx$ 7900  & 0.26      & $\approx$ 0.5\\
\cline{2-7}
   & 200 & 200      & 60.7  & $\approx$ 6000  & 5.4$\E{-2}$ & $\approx$
0.1\\
\cline{3-7}
   &     & $\infty$ & 49.4  & $\approx$ 5000  & 3.7$\E{-2}$ &  -\\
\hline
6  & 100 & 200      & 0.71  & 71              & 3.2$\E{-3}$ &  -\\
\cline{3-7}
   &     & $\infty$ & 0.38  & 38              & 2.1$\E{-3}$ &  -\\
\cline{2-7}
   & 200 & 200      & 0.16  & 16              & 1.7$\E{-4}$ &  -\\
\cline{3-7}
   &     & $\infty$ & 6.0$\E{-2}$ & 6         & 5.4$\E{-5}$ &  -\\
\hline
\end{tabular}
\caption{\label{wq-zahlen} 
        \em Cross section for charged Higgs
        pair production via gluon fusion in $pp$ ($p\bar{p}$) collisons for 
        hadron energies of $2$ and $14 \,\tev$, corresponding to the
        Tevatron and LHC. Additionally the number of 
        events in one year of running is given, assuming $2 \,\fb^{-1}$ 
        of integrated lumiosity for the Tevatron and $100 \,\fb^{-1}$ for 
        the LHC.
        }
\end{center}
\end{table}

Cross sections and numbers of events in table 
\ref{wq-zahlen} are given
for the projected collider energies  and integrated luminosities
of LHC and Tevatron Run II. 
At the Tevatron, the gluon fusion mechanism for charged Higgs pair
production is irrelevant.
Therefore, for the subsequent discussions we
will focus on the LHC with $\sqrt{S}=14$ TeV.

\begin{figure}[hbt]
\begin{center}
  \psfrag{XL}[l]{$\mu [\gev]$}
  \psfrag{YL}[b]{$\sigma [\fb]$}
  \psfrag{SIGMAPPGGHPHMX}[l]{$\sigma ( pp \to H^+ H^- +X)$}
  \psfrag{MSQ200}[l]{$\msqu = 200\, \gev$}
  \psfrag{TANB06X}[c]{$\scriptstyle \tan\beta = 6$}
  \psfrag{TANB1.5}[c]{$\scriptstyle \tan\beta = 1.5$}
  \psfrag{TANB50X}[c]{$\scriptstyle \tan\beta = 50$}
\resizebox*{1.0\width}{1.0\height}{\includegraphics*{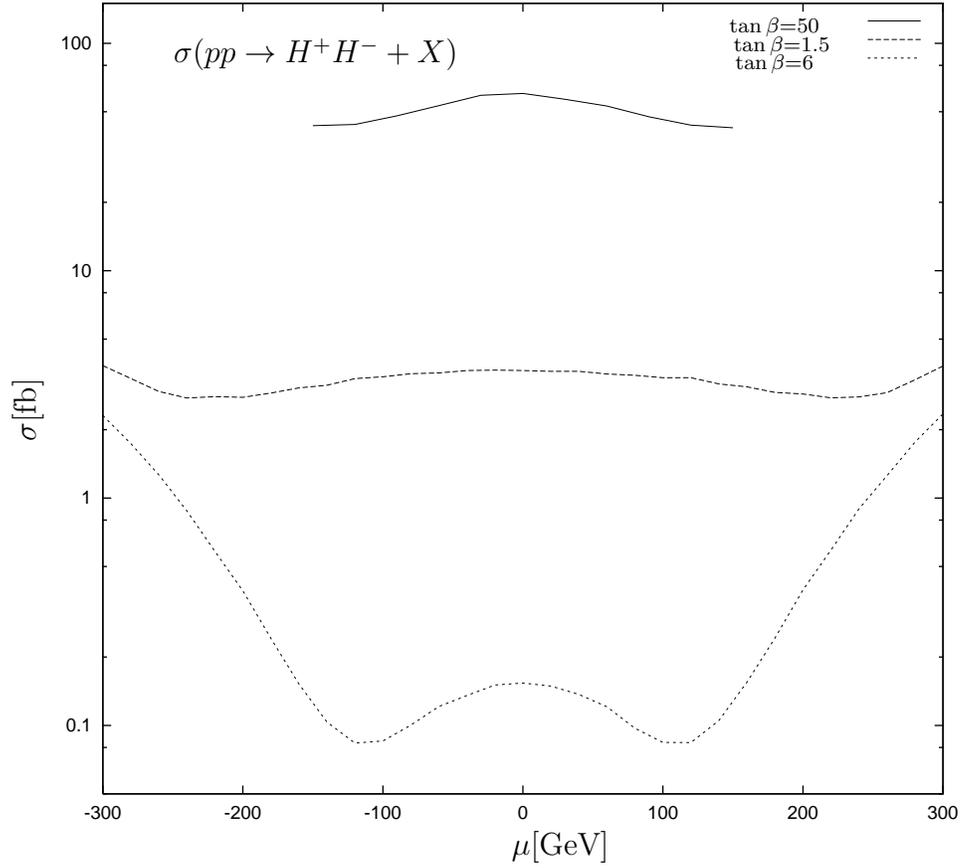}}
    \caption{\label{MU-depencence}
        \em $\mu$-dependence of the hadronic cross section 
	for $\mhpm = \msqu = 200\,\gev$ and 
        $A_t = 0$ and for three values of 
	$\tan\beta$ ($1.5/ 6 / 50$). For $\tan\beta = 50$ the graph
	is not extended over the full range because one of 
	the stop masses would become too light.}
  \end{center}
\end{figure}

\bigskip
The variation of the cross section with the MSSM parameter $\mu$
is displayed in Figure \ref{MU-depencence}. Except for $\tan\beta$ values
around the minimum of the cross section, the $\mu$-dependence
is very flat and the choice $\mu=0$ already illustrates the
typical size of the cross section very well. For $\tan\beta=1.5$
a sizeable increase occurs for larger
values of $|\mu|$ by an order of magnitude,
but still far below the high $\tan\beta$ scenario.

\subsection{Dependence on the Higgs parameters}\label{higgs-parameters}

\begin{figure}[htb]
\begin{center}
  \psfrag{XL}[l]{$\mhpm [\gev]$}
  \psfrag{YL}[b]{$\sigma [\fb]$}
  \psfrag{SIGMAPPGGHPHMX}[l]{$\sigma ( pp \to H^+ H^- + X)$}
  \psfrag{ALLDIAG}[cb]{\scriptsize komplett}
  \psfrag{QUARK}[cb]{\scriptsize nur Quarks}
  \psfrag{TANB6}{$\tan\beta = 6$}
  \psfrag{TANB1.5}{$\tan\beta = 1.5$}
  \psfrag{TANB50}{$\tan\beta = 50$}
  \resizebox*{1.0\width}{1.0\height}{\includegraphics*{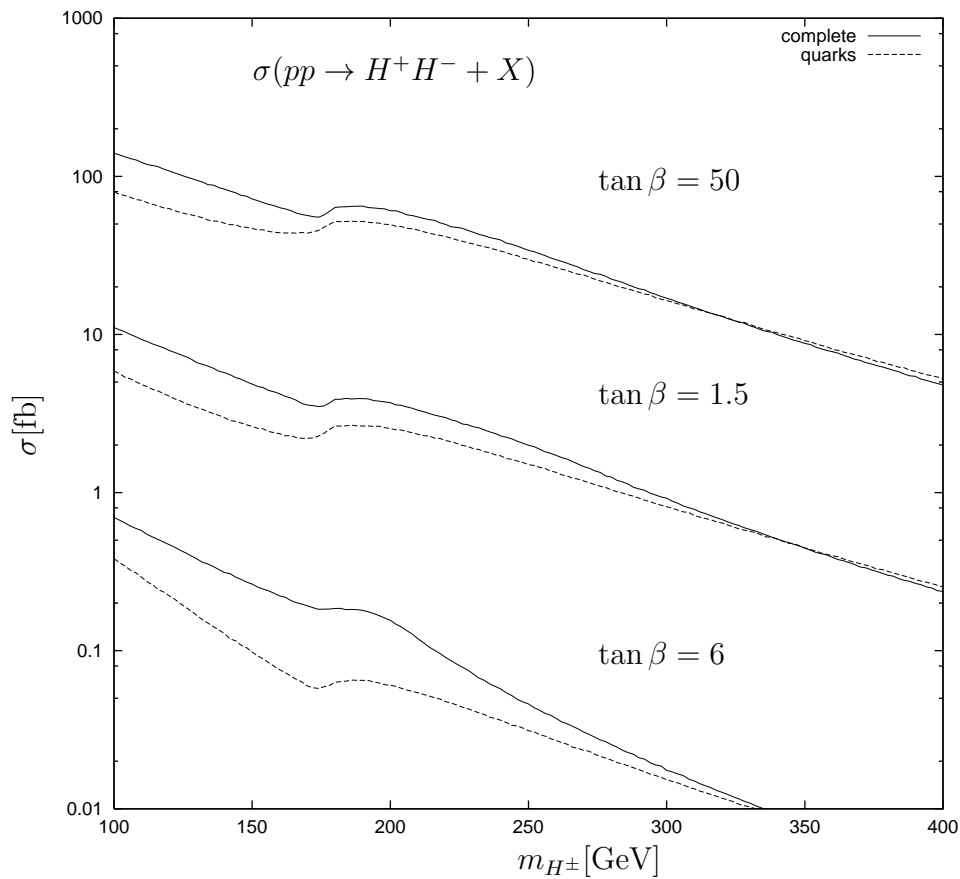}}
    \caption{\label{tb.mhpm.msq200} 
         \em Variation of the hadronic cross section (solid lines) with 
     	the mass of the charged Higgs particle for three values of
	$\tan\beta$ ($6/1.5/50$) and a squark mass scale of 
	$\msqu = 200\,\gev$. Additionally the contribution of only 
	the quark loops to the cross section is depicted (dashed lines), 
	which corresponds to the limit of large squark masses. 
	}          
  \end{center}
\end{figure}

The input parameters for the 
Higgs sector are $\mhpm$ and $\tan\beta$. 
Figure \ref{tb.mhpm.msq200} shows the dependence
of the hadronic cross section (\ref{hadronx})
 on the mass of the charged
Higgs for three representative values of $\tan\beta$ 
(including the minimum at $\tan\beta \approx 6$).  
The common squark
mass scale is $\msqu = 200\,\gev$ and all squark mixing is
switched off
($A_t = A_b = \mu = 0$). 
The hadronic cross section decreases almost exponentially with rising 
$\mhpm$. This behaviour is essentially  
explained by the decreasing number
of gluons at large values of $x$.
For Higgs masses near $m_t$, the top threshold gives rise to a step
with a slight increase of $\sigma$.

\smallskip
The dashed lines in Figure \ref{tb.mhpm.msq200}
show the cross section for decoupling
squarks.
The difference demonstrates the 
large effect squarks can have on the cross section whenever they are
not too heavy.
In the special case of $\mhpm = 170\,\gev$ and $\tan\beta = 6$ 
the cross section is more than 3 times the decoupling limit and
even for 
$\msqu = 500\,\gev$ the cross section is 
still 15 \% higher than in the decoupling case, independent of 
$\tan\beta$ and for charged Higgs masses in the range of Figure
\ref{tb.mhpm.msq200}.

\smallskip
The variation of the  hadronic cross section with $\tan\beta$
is displayed in Figure  \ref{tb.mhpm200.msq200.all-no3H}.
The cross section has a pronounced  minimum at
\begin{align*}
\tan\beta \approx \sqrt{\frac{m_t}{m_b}} \approx 6 \ehsp,
\end{align*}
which is a consequence of the
$\tan\beta$-dependence of the 
quark and squark couplings
to charged Higgs bosons.

\subsection{Dependence on the squark sector}\label{squark-contribution}

\begin{figure}[hbt]
\begin{center}   
  \psfrag{XL}[l]{$\msqu [\gev]$}
  \psfrag{YL}[b]{$\sigma [\fb]$}
  \psfrag{SIGMAPPGGHPHMX}[l]{$\sigma ( pp \to H^+ H^- +X),\,\tan\beta = 1.5$}
  \psfrag{MHPM200}{$\mhpm = 200\, \gev$}
  \psfrag{MHPM400}{$\mhpm = 400\, \gev$}
  \psfrag{MHPM600}{$\mhpm = 600\, \gev$}
  \psfrag{MHPM800}{$\mhpm = 800\, \gev$}
  \resizebox*{1.0\width}{1.0\height}{\includegraphics*{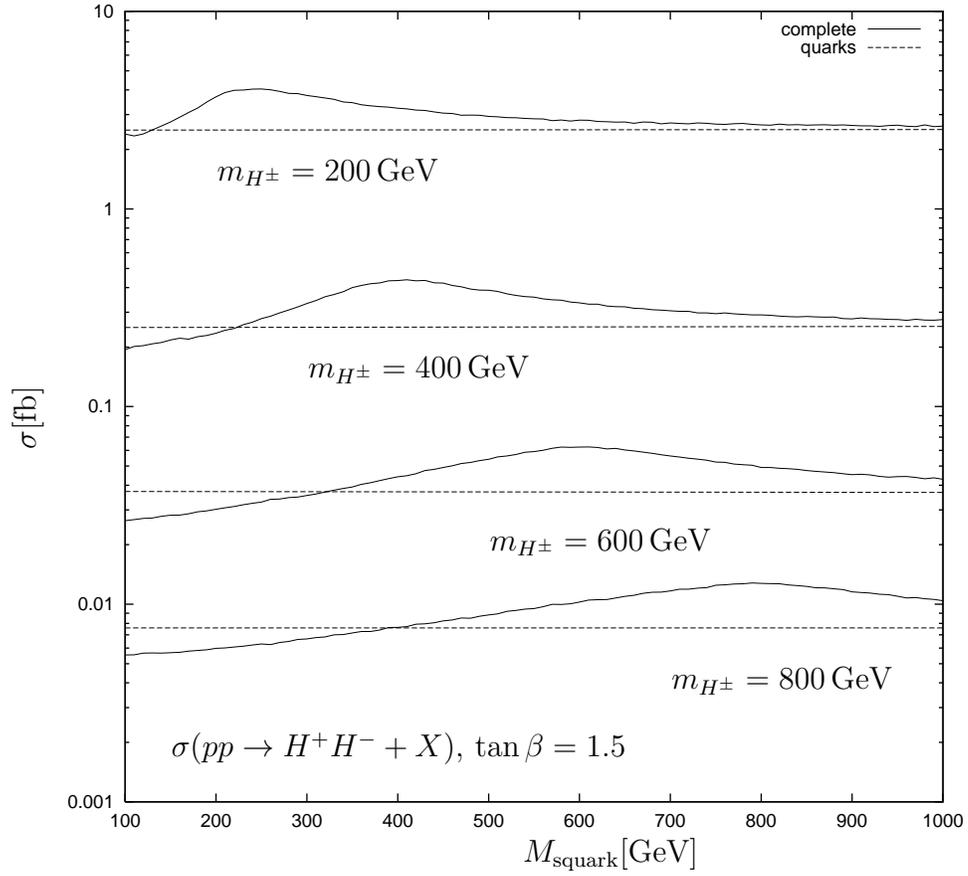}}
    \caption{\label{msq.multicompare}
      	\em Hadronic cross section for different values of $\mhpm$
	($200/400/600/800\,GeV$) versus the squark mass scale $\msqu$
	with $\tan\beta = 1.5$ (solid lines). For comparison the contribution 
	of only the quark loops is displayed in terms of the horizontal
	dashed lines.
        }
  \end{center}
\end{figure}

The contribution of the squark loop graphs (Appendix \ref{sq-graphs})
to the cross sections vanishes for large squark masses, due to the
decoupling property of the scalar quarks, and the pure fermion
loop result of \cite{KPSZ} is recovered
(see Figure \ref{msq.multicompare}).
The squark contribution is especially significant for squark masses
close to $\mhpm$, where the parton cross section near the
Higgs production threshold is enhanced by about a factor of 2 
compared to the cross section from quark loops only
(see Figure \ref{pwq.s}).
For squark masses lower than $\mhpm$, the 
squark loop amplitudes interfere mainly destructively with the quark loop 
amplitudes, thus diminishing the cross section 
as compared to the decoupling case.

\smallskip
The predominant squark contribution comes from the third generation
with their large Yukawa couplings. The structure of these couplings gives
rise to the minimum behaviour of $\sigma$ as a function of
$\tan\beta$. There are, however, also squark couplings 
independent of the quark masses and hence not suppressed by light quark
masses 
for the two other generations 
(see Appendix \ref{couplings}).
These contributions,
although much smaller in size,
are not completely negligible: for $\tan\beta\sim 6$
they constitute 45\% of the cross section. Since they are
essentially $\tan\beta$ independent, their relative influence
decreases at other values of $\tan\beta$ with larger cross sections. 

\begin{figure}[hbt]
\begin{center}   
  \psfrag{XL}[l]{$A_t [\gev]$}
  \psfrag{YL}[b]{$\sigma [\fb]$}
  \psfrag{SIGMAPPGGHPHMX}[l]{$\sigma ( pp \to H^+ H^- +X)$}
  \psfrag{MSQ200}[l]{$\msqu = 200\, \gev$}
  \psfrag{TANB033}[r]{$\scriptstyle \tan\beta = 33$}
  \psfrag{TANB108}[r]{$\scriptstyle \tan\beta = 1.08$}
  \resizebox*{1.0\width}{1.0\height}{\includegraphics*{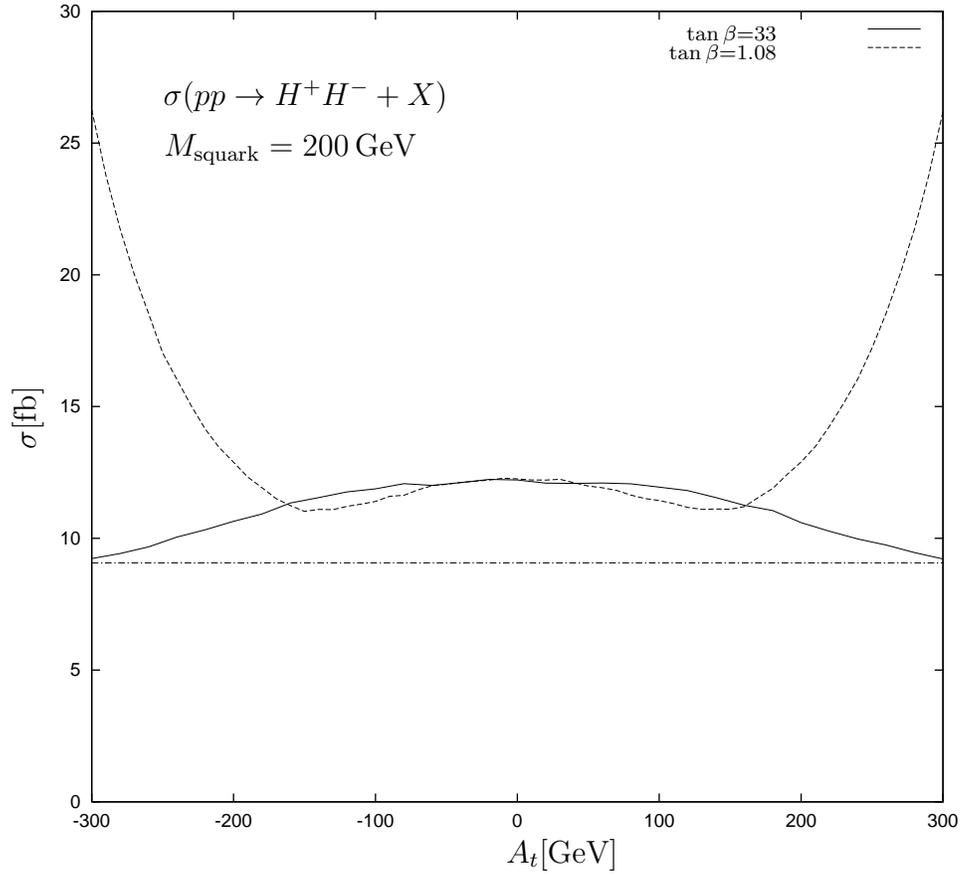}}
    \caption{\label{sq-mixing.At}
      	\em Hadronic cross section for $\mhpm = 200\,\gev$ and 
	two values of $\tan\beta$ ($33$: solid, $1.08$: dashed) versus 
	the mixing parameter $A_t$. For comparison the dot-dashed 
	line shows the cross section for $\tan\beta = 1.08$ 
	with the pure fermion loop contribution.
        }
  \end{center}
\end{figure}

\bigskip
The effects of
mixing in the stop sector are visualized in Figure 
\ref{sq-mixing.At}, where the non-diagonal entry $A_t$
in the stop mass matrix is varied (for $\mu=0$),
for two $\tan\beta$-values (1.08 and 33).
For small values of $A_t$, the cross sections are nearly
identical. In the low $\tan\beta$ case, however, 
a significantly higher 
cross section can be obtained by increasing the value of $A_t$,
with a strong dependence on $A_t$.

\subsection{Sensitivity to the Higgs self interaction}

Analogously to the standard Higgs case 
\cite{GB} one may ask wether charged 
Higgs pair production is sensitive to the Higgs self interaction
that enters through the neutral Higgs exchange graphs 
(see Appendix \ref{sq-graphs}). 
The situation here is less transparent
than in the standard model
because of the two neutral Higgs particles 
with different couplings to the charged Higgs pair.

\begin{figure}[hbt]
\begin{center}
  \psfrag{XL}[l]{$\mhpm [\gev]$}
  \psfrag{YL}[b]{$\sigma [\fb]$}
  \psfrag{TANB6}{$\tan\beta = 6$}
  \psfrag{TANB1.5}{$\tan\beta = 1.5$}
  \psfrag{TANB50}{$\tan\beta = 50$}
  \psfrag{SIGMAPPGGHPHMX}[l]{$\sigma ( pp \to H^+ H^- + X)$}
  \resizebox*{1.0\width}{1.0\height}{\includegraphics*{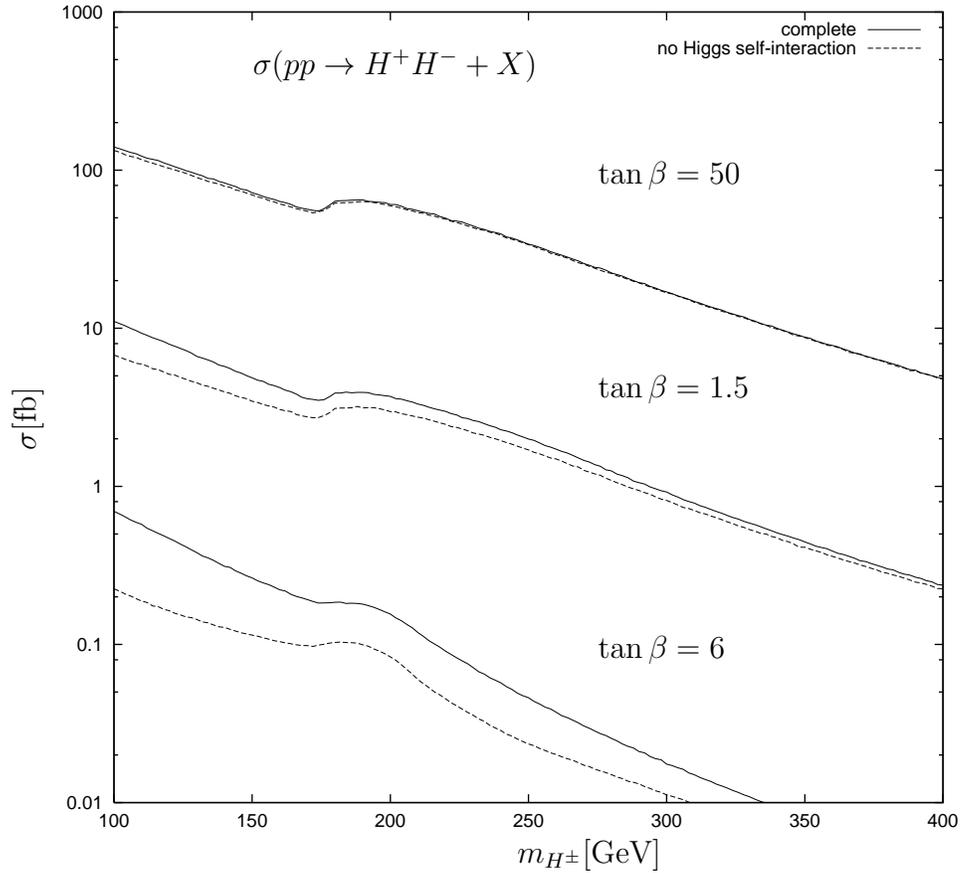}}
    \caption{ \label{no3h.msq200}
         \em Comparison between the hadronic cross section versus the
	charged Higgs mass with and without
	the trilinear Higgs coupling (solid and dashed line, respectively) 
	for three values of $\tan\beta$ ($6/1.5/50$). The 
	squark mass scale is $\msqu = 200\,\gev$.
	} 
  \end{center}
\end{figure}

To illustrate the influence of the Higgs self interaction, 
Figures \ref{no3h.msq200} and \ref{tb.mhpm200.msq200.all-no3H} 
show the complete cross section (solid lines) versus the cross section
evaluated with the Higgs self interaction turned off (dashed lines).
In both Figures the 
same squark mass scenario  as in chapter 
\ref{higgs-parameters} has been adopted 
(i.e. $\msqu = 200 \:\gev$ and no squark mixing).
Except around the minimum, the relative contribution of the Higgs 
self interaction is much smaller than the contributions from the
box diagrams
(the observation of the smallness of the diagrams with the quark/squark
triangles has already been made in \cite{willenbrock}).

\section{Conclusions}

Gluon-gluon 
fusion is an important mechanism for producing charged Higgs particles 
at the LHC.
Although at smaller rates than in $q\bar{q}$ annihilation,
it is sensitive also to the scalar quark sector if the Higgs
bosons are supersymmetric.   For not too large masses,
virtual squarks are not negligible and can increase the
cross section up to a factor 2, in special cases even 3, compared 
to the decoupling limit
with heavy squarks. 
For squark masses not smaller  than $\mhpm$ the squark contribution
is always positive. 
An interesting behaviour is observed for squark mixing:
For low values of $\tan\beta$ the cross section
is strongly dependent on the trilinear top squark coupling $A_t$,
yielding a steep rise for increasing values of $A_t$.
Charged Higgs boson pairs can thus provide a valuable source of
information, useful 
to disentangle various classes of models. 

\bigskip\bigskip
\noindent {\em Acknowledgement.} We would like to thank A.~Barrientos and 
B.~Kniehl for helpful discussions.


\appendix

\section{Tensor basis}\label{tensorbasis}

The amplitude $\MM^{fi}_{\lambda \lambda'}$ for the fusion of two
massless vector particles into two scalar particles is decomposed as
follows:
\begin{equation*}
\MM^{fi}_{\lambda \lambda'}=
        N_{fi}\:
        \polv{\lambda}{\mu}{p_1}\polv{\lambda'}{\nu}{p_2}
        \widetilde{\MM}_{\mu\nu} \ehsp, 
        \ehsp \widetilde{\MM}_{\mu\nu} = \sum_{i=1}^{4} F_i A_i^{\mu\nu}
        \ehsp,
\end{equation*}
with a normalization factor $N_{fi}$ and the following tensor basis
\cite{GB,PSZ}:
\begin{align}
A_1^{\mu\nu} & = g^{\mu\nu}-\frac{p_2^\mu p_1^\nu}{p_1p_2}\\
A_2^{\mu\nu} & = g^{\mu\nu}+\frac{p_3^2 p_2^\mu p_1^\nu}{p_t^2(p_1p_2)}
        -\frac{2(p_2p_3) p_3^\mu p_1^\nu}{p_t^2(p_1p_2)}
        -\frac{2(p_1p_3) p_2^\mu p_3^\nu}{p_t^2(p_1p_2)}
        +\frac{2 p_3^\mu p_3^\nu}{p_t^2}\\
A_3^{\mu\nu} & = \frac{\epsilon^{\mu\nu p_1 p_2}}{(p_1p_2)}\\
A_4^{\mu\nu} & = \frac{ p_3^\mu \epsilon^{\nu p_1 p_2 p_3} 
                +p_3^\nu \epsilon^{\mu p_1 p_2 p_3} 
        +(p_2p_3)\epsilon^{\mu\nu p_1 p_3}
                +(p_1p_3)\epsilon^{\mu\nu p_2 p_3} }{p_t^2(p_1p_2)}
\ehsp.
\end{align}
The transverse momentum squared $p_t^2$ is defined as
\begin{align*}
p_t^2 & = 2\frac{(p_1p_3)(p_2p_3)}{(p_1p_2)}-p_3^2
\ehsp.
\end{align*}
The properties of these tensors, which have been exploited in the calculation,
are tranversality:
\begin{align*}
  p_{1,\mu} A_i^{\mu\nu} = 0\, , \quad & p_{2,\nu} A_i^{\mu\nu} = 0
\end{align*}
and orthogonality:
\begin{align*}
  A_i^{\mu\nu} A_{j,\mu\nu} & = 2\: \delta_{ij}
  \ehsp.
\end{align*}

\section{Form factors}\label{formfactors}

For 3- and 4-point functions the following shorthand notations are defined:
\begin{align*}
C^{ABC}_{lm}   & = \frac{1}{i \pi^2}
	\idvk \frac{1}{
	[k^2\!-\!m_A^2][(k\!+\!p_l)^2\!-\!m_B^2][(k\!+\!p_l\!+\!p_m)^2\!-\!m_C^2]} \ehsp,\\
D^{ABCD}_{klm} & = \frac{1}{i \pi^2}
        \idvk \frac{1}{
        [k^2\!-\!m_A^2][(k\!+\!p_k)^2\!-\!m_B^2]
	[(k\!+\!p_k\!+\!p_l)^2\!-\!m_C^2][(k\!+\!p_k\!+\!p_l\!+\!p_m)^2\!-\!m_D^2]} \ehsp.
\end{align*}

\subsection{Squark contributions}

The form factors from squark loops with two or three internal
lines are:
\begin{align}
F_\Sq^{\circ} + F_{\Sq}^\triangle  & = \frac{1}{4\pi^2}
\sum_{l,k=1}^{2} 
\Bigg[
 f_{\Sq,1}^{\circ+\triangle} (l,k)
+f_{\Sq,2}^{\circ+\triangle} (l,k)
\Bigg]
\end{align}
with 
\begin{align}
f_{\Sq,1}^{\circ+\triangle} (l,k) & =
\sum_{\Sq = \Stop,\Sbot} 
\Bigg[ \delta_{lk} \,
	\Big(
		\frac{1}{2} + m_{\Sq_l}^2 C^{\Sq_l\Sq_l\Sq_l}_{12} 
	\Big)
	\Big(
		\gsqlsqlhphm
	       - \sum_{i=1}^{2} 
		 \frac{\ghihphm\: \gsqlsqlhi}
		      { s-m_{H^0_i}^2 }
	\Big)
\Bigg]\\
f_{\Sq,2}^{\circ+\triangle} (l,k) & = 
	\frac{3}{4} V_{lk} \:
	\Big(                               
	C^{\Stop_k\Sbot_l\Stop_k}_{34} + C^{\Sbot_l\Stop_k\Sbot_l}_{34}
	\Big)
\ehsp,
\end{align}
and
\begin{align}
G_\Sq^{\circ}+G_\Sq^{\triangle} = \frac{1}{4\pi^2}
\sum_{l,k=1}^{2} V_{lk}
\Bigg[ \frac{1}{4} \big( 2 + \frac{\mhpm^2}{p_t^2} \big)
\Big( C^{\Stop_k\Sbot_l\Stop_k}_{34} + C^{\Sbot_l\Stop_k\Sbot_l}_{34} \Big)
\Bigg] \ehsp. 
\end{align}
$V_{lk}$ is a shorthand notation for the product 
of two couplings of squarks to
the charged Higgs particle which appears again in the box amplitudes:
\begin{align}
V_{lk} = \ghpmstopsbottom{k}{l}^2 \ehsp.
\end{align}
The form factors originating from squark box diagramms are:
\begin{align}
F_\Sq^{\square} = \frac{1}{4\pi^2} \sum_{l,k=1}^{2} \; V_{lk} 
\Bigg[ 
  \big(
  f_\Sq^\square (\Stop_k,\Sbot_l ; \hat{t},\hat{u})
+ f_\Sq^\square (\Sbot_l,\Stop_k ; \hat{t},\hat{u})
  \big)
+ \big( \hat{t} \leftrightarrow \hat{u} \big)
\Bigg]
\end{align}
with
\begin{multline}
f_\Sq^\square (\Stop_k,\Sbot_l ; \hat{t},\hat{u}) = 
\frac{(\hat{t}-\mhpm^2)}{2\,\hat{s}} C^{\Stop_k\Stop_k\Sbot_l}_{13} 
     -\frac{3}{8} C^{\Stop_k\Sbot_l\Stop_k}_{34}
     -\frac{m_{\Stop_k}^2}{2} D^{\Stop_k\Stop_k\Stop_k\Sbot_l}_{123} 
     -\frac{m_{\Stop_k}^2 + m_{\Sbot_l}^2 + p_t^2}{8} 
      D^{\Stop_k\Stop_k\Sbot_l\Sbot_l}_{132}
\ehsp,
\end{multline}
and
\begin{align}
G_\Sq^{\square} = \frac{1}{4\pi^2} \sum_{l,k=1}^{2} \; V_{lk}
\Bigg[
  \big(
  g_\Sq^{\square} (\Stop_k,\Sbot_l ; \hat{t},\hat{u}) 
+ g_\Sq^{\square} (\Sbot_l,\Stop_k ; \hat{t},\hat{u})
  \big)
+ \big( \hat{t} \leftrightarrow \hat{u} \big)
\Bigg]
\end{align}
with
\begin{multline}
g_\Sq^{\square} (\Stop_k,\Sbot_l ; \hat{t},\hat{u}) = \frac{1}{2 p_t^2}
\Bigg[
\frac{1}{2}
\big( m_{\Stop_k}^2 - m_{\Sbot_l}^2+\mhpm^2 - \frac{\hat{s}}{2} \big)
     C^{\Stop_k\Stop_k\Stop_k}_{12}
    +\frac{\hat{t}\, (\hat{t}-\mhpm^2)}{\hat{s}} C^{\Stop_k\Stop_k\Sbot_l}_{13}\\
    +\frac{1}{4} (3 \mhpm^2 - \hat{s}) C^{\Stop_k\Sbot_l\Stop_k}_{34}
    -\frac{1}{4} 
               \big( 
                 ( m_{\Stop_k}^2 + m_{\Sbot_l}^2 ) p_t^2 
                +( m_{\Stop_k}^2 - m_{\Sbot_l}^2 )^2 
               \big) D^{\Stop_k\Stop_k\Sbot_l\Sbot_l}_{132}\\           
             -\frac{1}{2} \big(
                 (\hat{t} - m_{\Sbot_l}^2 )^2
                +m_{\Stop_k}^2 (m_{\Stop_k}^2 -2 m_{\Sbot_l}^2 -2\frac{(\hat{t}-\mhpm^2)^2}{\hat{s}})
               \big) D^{\Stop_k\Stop_k\Stop_k\Sbot_l}_{213} 
\Bigg] \ehsp.
\end{multline}

\subsection{Quark contributions}

The form factor due to triangle quark loops is:
\begin{equation}
F_{\text{q}}^\triangle = \frac{1}{4\pi^2}
\sum_{i=1}^{2}
\Bigg[
  f_{\text{q}}^\triangle (H^0_i,t) 
+ f_{\text{q}}^\triangle (H^0_i,b) 
\Bigg]
\end{equation}
with
\begin{equation}
f_{\text{q}}^\triangle (H^0_i,q) = 
\frac{\ghiqq \: \ghihphm}{ \hat{s}-m_{H^0_i}^2 +i m_{H^0_i}\Gamma_{H^0_i} }
m_q \Big( 2 + (4 m_q^2 - \hat{s}) C^{qqq}_{12} \Big) \ehsp,
\end{equation}
and the form factors from quark box graphs are: 
\begin{multline}
F_{\text{q}}^\square = \frac{1}{4\pi^2} 
\Bigg[ \bigg(
    {\gshpmtb}^2 \Big( f_{\text{q}}^\square (m_t,m_b ; \hat{t},\hat{u})  
		     + f_{\text{q}}^\square (m_b,m_t ; \hat{t},\hat{u})  \Big)\\
  + {\gphpmtb}^2 \Big( f_{\text{q}}^\square (m_t,-m_b ; \hat{t},\hat{u}) 
		     + f_{\text{q}}^\square (m_b,-m_t ; \hat{t},\hat{u}) \Big) 
       \bigg) 
+ \bigg( \hat{t} \leftrightarrow \hat{u} \bigg) 
\Bigg]
\end{multline}
with 
\begin{multline}
f_{\text{q}}^\square ( m_t , m_b ; \hat{t},\hat{u}) =
\frac{1}{2} + m_t^2 C^{ttt}_{12} +
\left( \mhpm^2 - ( m_t + m_b )^2 \right) \frac{(\hat{t}-\mhpm^2)}{\hat{s}} C^{ttb}_{13}
- m_t m_b \frac{\hat{s}}{4} ( 2 D^{tttb}_{123} + D^{ttbb}_{132} )\\
- \frac{1}{4}\left( \mhpm^2 - ( m_t + m_b )^2 \right) 
\Big( ( p_t^2 + m_b^2 + m_t^2 ) D^{ttbb}_{132} 
+ 4 m_t^2 D^{tttb}_{123} \Big)
\ehsp;
\end{multline}
\begin{multline}
G_{\text{q}}^\square = \frac{1}{4\pi^2}
\Bigg[ 
	\bigg(
    {\gshpmtb}^2 \Big( g_{\text{q}}^\square (m_t,m_b ; \hat{t},\hat{u})  
		     + g_{\text{q}}^\square (m_b,m_t ; \hat{t},\hat{u})  \Big)\\
  + {\gphpmtb}^2 \Big( g_{\text{q}}^\square (m_t,-m_b ; \hat{t},\hat{u}) 
	             + g_{\text{q}}^\square (m_b,-m_t ; \hat{t},\hat{u}) \Big)
	\bigg)
+ \bigg( \hat{t} \leftrightarrow \hat{u} \bigg)
\Bigg]
\end{multline}
with
\begin{multline}
g_{\text{q}}^\square ( m_t , m_b ; \hat{t},\hat{u} ) = \frac{1}{4 p_t^2} 
\Bigg[
\big( \frac{\hat{s}}{2} + ( m_t + m_b )^2 - \mhpm^2 \big) \Big(
   \big( 2 p_t^2 + \frac{(\hat{u}-\hat{t})^2}{\hat{s}} \big) C^{tbt}_{34}\\
   +\big( 2 ( \mhpm^2 + m_t^2 - m_b^2 ) - \hat{s} \big) C^{ttt}_{12} \Big)
- \hat{s} p_t^2 C^{ttt}_{12}
+2 \frac{(\hat{t}-\mhpm^2)}{\hat{s}} \Big( \mhpm^4 + \hat{t}^2 - 2 \hat{t} ( m_t + m_b )^2 \Big) C^{ttb}_{13}\\
+p_t^2 \Big(
        \big( \hat{s} m_b^2 - 2 m_t^2 ( 2 m_b^2 -m_t^2 +\mhpm^2) \big)  D^{tttb}_{123} 
       +\big( \frac{\hat{s}}{2} ( m_t^2 + m_b^2 ) + P_1(m_t,m_b,\mhpm) \big) D^{ttbb}_{132} \Big)\\
+ m_t^4\frac{(\hat{t}-\mhpm^2)^2}{\hat{s}} \Big( D^{tttb}_{123} - 3 D^{tttb}_{213} \Big)
-\frac{\hat{u}}{\hat{s}} m_t^2 (\hat{t}^2 - \mhpm^4) ( D^{tttb}_{123} + D^{tttb}_{213} )\\ 
-2 \big(   
    m_b^2 m_t^2 \frac{(\hat{t}-\hat{u})^2}{\hat{s}} - P_2(m_t,m_b,\mhpm)
   \big) D^{tttb}_{123}\\
+\Big( -\hat{t}^3 + 4 \hat{t}^2 m_b ( m_t + m_b) 
	- \hat{t} (8 m_b^3 m_t + 4 m_b^4 +\mhpm^4)\\
+8 m_b m_t^3 \big( (\hat{t}-\hat{u}) -\frac{(\hat{u}-\mhpm^2)^2}{\hat{s}} \big)
      +(\hat{s} - 2 \mhpm^2) m_b^4 \Big) D^{tttb}_{213}\\
+\frac{1}{2} ( m_t^2 - m_b^2 )^2 \Big(
2 ( \mhpm^2 + ( m_t + m_b )^2 ) + \frac{(\hat{u}-\hat{t})^2}{\hat{s}} \Big) D^{ttbb}_{132}
\Bigg]
\end{multline}
containing the polynomials
\begin{align}
P_1(m_t,m_b,\mhpm) &  = 3 m_t^4 + 2 m_t^3 m_b - 2 m_t^2 m_b^2 + 2 m_t m_b^3 
                   + 3 m_b^4 - \mhpm^2 (m_t^2+m_b^2)\\[.3cm]
\nonumber
P_2(m_t,m_b,\mhpm) &  = m_t^6 + 2 m_t^5 m_b  - m_t^4 m_b^2 - 4 m_t^3 m_b^3 
                   - m_t^2 m_b^4 + 2 m_t m_b^5 + m_b^6\\
               &  - \mhpm^4 (m_t^2-m_b^2) - 2 \mhpm^2 m_t^2 m_b^2
\ehsp;
\end{align}
\begin{equation}\label{Hqbox}
H_{\text{q}}^\square = \frac{1}{4\pi^2} \gshpmtb \gphpmtb 
\Big[ \big( 
      h_{\text{q}}^\square(m_t,m_b ; \hat{t},\hat{u}) 
    + h_{\text{q}}^\square(m_b,m_t ; \hat{t},\hat{u}) 
      \big)
- \big( \hat{t} \leftrightarrow \hat{u} \big) 
\Big]
\end{equation}
with 
\begin{multline}
h_{\text{q}}^\square(m_t,m_b ; \hat{t},\hat{u}) = \frac{1}{2 p_t^2}
\Bigg[
 \hat{t} \Big( 
	( \hat{s} -2 \mhpm^2 +2 m_b^2- 2 m_t^2 ) C^{ttt}_{12} 
        +(\hat{s}-4 \mhpm^2) C^{tbt}_{34} 
	\Big)\\
-2 \frac{(\hat{t}-\mhpm^2)^2}{\hat{s}} (\hat{t}+\mhpm^2) C^{ttb}_{13}
+p_t^2 \hat{t} \Big(
	 m_t^2 (D^{tttb}_{123}+D^{tttb}_{213})
	+(m_t^2+m_b^2) D^{ttbb}_{132} 
	\Big)\\
+ \hat{t} (m_t^2-m_b^2)^2 \Big( 
	D^{tttb}_{123}+D^{tttb}_{213}+D^{ttbb}_{132} 
	\Big)
+\Big( 
	2 \frac{(\hat{u}-\mhpm^2)}{\hat{s}} (\hat{t}-\hat{u}) m_t^2 \mhpm^2 \\
	+ (\hat{u}^2 - \mhpm^4) (2 \frac{(\hat{u}-\mhpm^2)}{\hat{s}} m_t^2-\hat{u}+2 m_b^2) -\hat{s} p_t^2 m_b^2 
	\Big) D^{tttb}_{123} \Bigg] \ehsp.
\end{multline}


\section{MSSM couplings}\label{couplings}

In the following all couplings of third generation 
quarks and squarks to MSSM Higgs particles, that 
are relevant to the process, are collected. 
The factor $g_2 = e/s_w$ denotes the coupling constant of the 
weak interaction, $s_w = \sin\theta_w$, $c_w = \cos\theta_w$
and $t_w = \tan\theta_w$ with the weak mixing angle $\theta_w$.
The subscripts 's' and 'p' in 
$g_s[\cdots]$ and $g_p[\cdots]$ distinguish between
the scalar and pseudoscalar couplings in the Higgs--fermion interactions.

\subsection{Higgs self couplings}

\begin{equation}
g [H^0_1, H^{+}, H^{-}]=
        -g_2 
        \left[
        m_W\cos(\beta-\alpha) 
      - \frac{m_Z}{2 c_w}\cos 2\beta\cos(\alpha+\beta)
        \right]
\ehsp,
\end{equation}
\begin{equation}
g [H^0_2, H^{+}, H^{-}]=
        -g_2
        \left[
        m_W\sin(\beta-\alpha) 
      + \frac{m_Z}{2 c_w}\cos 2\beta\sin(\alpha+\beta)
        \right]
\ehsp.
\end{equation}

\subsection{Quark couplings to Higgs bosons}

\subsubsection*{Neutral Higgs bosons} 

\begin{align}
g_s[H^{0}_{1}, t, t ] & =
        -g_2\:\frac{m_t}{2 m_W}\:\frac{\sin\alpha}{\sin\beta} 
\ehsp,
&  
g_s[H^{0}_{1}, b, b ] & =
        -g_2\:\frac{m_b}{2m_W}\:\frac{\cos\alpha}{\cos\beta}
\ehsp,\\
g_s[H^{0}_{2}, t, t ] & =
        -g_2\:\frac{m_t}{2m_W}\:\frac{\cos\alpha}{\sin\beta}
\ehsp,
&
g_s[H^{0}_{2}, b, b ] & =
        +g_2\:\frac{m_b}{2m_W}\:\frac{\sin\alpha}{\cos\beta}
\ehsp,\\
g_p[H^{0}_{i}, t, t ] & = 0
\ehsp,
&
g_p[H^{0}_{i}, b, b ] & = 0
\ehsp.
\end{align}

\subsubsection*{Charged Higgs bosons}

\begin{align}
g_s[H^+\:\text{out}, t\:\text{in}, b\:\text{out}] & =
        g_2\,\frac{m_b\tan\beta + m_t\cot\beta}{2\sqrt{2} m_W} =
        g_s[H^-\:\text{out}, b\:\text{in}, t\:\text{out}] \\
g_p[H^+\:\text{out}, t\:\text{in}, b\:\text{out}] & =
        -g_2\,\frac{m_b\tan\beta - m_t\cot\beta}{2\sqrt{2} m_W} =
        -g_p[H^-\:\text{out}, b\:\text{in}, t\:\text{out}]
\end{align}

\subsection{Squark couplings to Higgs bosons}

\subsubsection*{Neutral Higgs bosons}

\begin{multline}
g[H^0_1,\Stop_1,\Stop_1] = g_2\, 
\Big[  \frac{m_t}{2 m_W \sin\beta} 
	\Big( (A_t \sin\alpha - \mu\cos\alpha) \sin 2\theta_{\Stop}
		-2 m_t \sin\alpha 
	\Big) \\
      + \frac{m_Z \cos(\alpha+\beta)}{6 c_w} 
	\Big( (5 - 8 c_w^2) \cos^2\theta_{\Stop} -4 s_w^2
        \Big)	
\Big]
\end{multline}

\begin{multline}
g[H^0_1,\Stop_2,\Stop_2] = g_2\, 
\Big[  \frac{m_t}{2 m_W \sin\beta} 
	\Big( - (A_t \sin\alpha - \mu\cos\alpha) \sin 2\theta_{\Stop}
		-2 m_t \sin\alpha 
	\Big) \\
      + \frac{m_Z \cos(\alpha+\beta)}{6 c_w} 
	\Big( - (5 - 8 c_w^2) \cos^2\theta_{\Stop} + (1 -4 s_w^2)
        \Big)	
\Big]
\end{multline}

\begin{multline}
g[H^0_1,\Sbot_1,\Sbot_1] = g_2\, 
\Big[  \frac{m_b}{2 m_W \cos\beta} 
	\Big( (A_b \cos\alpha -\mu\sin\alpha ) \sin 2\theta_{\Sbot}
		-2 m_b \cos\alpha
	\Big) \\
      + \frac{m_Z \cos(\alpha+\beta)}{6 c_w} 
	\Big( (4 c_w^2 - 1 ) \cos^2\theta_{\Sbot} + 2 s_w^2
        \Big)	
\Big]
\end{multline}

\begin{multline}
g[H^0_1,\Sbot_2,\Sbot_2] = g_2\, 
\Big[  \frac{m_b}{2 m_W \cos\beta} 
	\Big( - (A_b \cos\alpha -\mu\sin\alpha ) \sin 2\theta_{\Sbot}
		-2 m_b \cos\alpha
	\Big) \\
      + \frac{m_Z \cos(\alpha+\beta)}{6 c_w} 
	\Big( - (4 c_w^2 - 1 ) \cos^2\theta_{\Sbot} + (1 + 2 c_w^2)
        \Big)	
\Big]
\end{multline}

\begin{align}
g[H^0_2,\Sq_i,\Sq_i] & = g[H^0_1,\Sq_i,\Sq_i] \:
\big( \sin\alpha \to \cos\alpha\, ,\: \cos\alpha \to -\sin\alpha 
\big)
\end{align}

\subsubsection*{Charged Higgs bosons}

\begin{multline}
g[ H^+\:\text{out}, \Stop_1\:\text{in}, \Sbot_1\:\text{out} ]  = 
g[ H^-\:\text{out}, \Stop_1\:\text{out}, \Sbot_1\:\text{in} ]  =\\
\frac{ { g_2} }{2 \sqrt{2} { m_W} } \Big[
	\sin 2 \beta \Big( m_b^2 (1 + \tan^2\beta )
                         + m_t^2 (1 + \cot^2\beta ) - 2 m_W^2 \Big) 
	\:\cos\theta_{\Sbot} \cos\theta_{\Stop}\\
+ 2 m_t m_b ( \tan\beta + \cot\beta ) \: \sin\theta_{\Sbot} \sin\theta_{\Stop}
- 2 m_b ( \mu + A_b \tan\beta ) \: \sin\theta_{\Sbot} \cos\theta_{\Stop}\\
- 2 m_t ( \mu + A_t \cot\beta ) \: \cos\theta_{\Sbot} \sin\theta_{\Stop}
\Big]
\end{multline}

\begin{multline}
g[ H^+\:\text{out} ,\Stop_1\:\text{in} ,\Sbot_2\:\text{out}  ] = 
g[ H^-\:\text{out} ,\Stop_1\:\text{out} ,\Sbot_2\:\text{in}  ] =\\
\frac{ { g_2} }{2 \sqrt{2} { m_W} } \Big[
	\sin 2 \beta \Big( m_b^2 (1 + \tan^2\beta ) 
                         + m_t^2 (1 + \cot^2\beta ) - 2 m_W^2 \Big) 
	\:\sin\theta_{\Sbot} \cos\theta_{\Stop}\\
- 2 m_t m_b ( \tan\beta + \cot\beta ) \: \cos\theta_{\Sbot} \sin\theta_{\Stop}
+ 2 m_b ( \mu + A_b \tan\beta ) \: \cos\theta_{\Sbot} \cos\theta_{\Stop}\\
- 2 m_t ( \mu + A_t \cot\beta ) \: \sin\theta_{\Sbot} \sin\theta_{\Stop}
\Big]
\end{multline}

\begin{multline}
g[H^+\:\text{out} ,\Stop_2\:\text{in} ,\Sbot_1\:\text{out} ] = 
g[H^-\:\text{out} ,\Stop_2\:\text{out} ,\Sbot_1\:\text{in} ] =\\
\frac{ { g_2} }{2 \sqrt{2} { m_W} } \Big[
	\sin 2 \beta \Big( m_b^2 (1 + \tan^2\beta ) 
                         + m_t^2 (1 + \cot^2\beta ) - 2 m_W^2 \Big) 
	\:\cos\theta_{\Sbot} \sin\theta_{\Stop}\\
- 2 m_t m_b ( \tan\beta + \cot\beta ) \: \sin\theta_{\Sbot} \cos\theta_{\Stop}
+ 2 m_b ( \mu + A_b \tan\beta ) \: \sin\theta_{\Sbot} \sin\theta_{\Stop}\\
+ 2 m_t ( \mu + A_t \cot\beta ) \: \cos\theta_{\Sbot} \cos\theta_{\Stop}
\Big]
\end{multline}

\begin{multline}
g[ H^+\:\text{out} ,\Stop_2\:\text{in} ,\Sbot_2\:\text{out}  ] = 
g[ H^-\:\text{out} ,\Stop_2\:\text{out} ,\Sbot_2\:\text{} in ] =\\
\frac{ { g_2} }{2 \sqrt{2} { m_W} } \Big[
	\sin 2 \beta \Big( m_b^2 (1 + \tan^2\beta ) 
                         + m_t^2 (1 + \cot^2\beta ) - 2 m_W^2 \Big) 
	\:\sin\theta_{\Sbot} \sin\theta_{\Stop}\\
+ 2 m_t m_b ( \tan\beta + \cot\beta ) \: \cos\theta_{\Sbot} \cos\theta_{\Stop}
+ 2 m_b ( \mu + A_b \tan\beta ) \: \cos\theta_{\Sbot} \sin\theta_{\Stop}\\
+ 2 m_t ( \mu + A_t \cot\beta ) \: \sin\theta_{\Sbot} \cos\theta_{\Stop}
\Big]
\end{multline}

\begin{multline}
g[\Stop_1,\Stop_1,H^+,H^-] = \frac{ g_2^2}{2} 
\Big[\frac{1}{3} t_w^2 \cos 2\beta - \frac{m_t^2}{m_W^2} \cot^2\beta\\
- \cos^2\theta_{\Stop} 
	\Big( (\frac{1}{2} + \frac{5}{6} t_w^2) \cos 2\beta 
	+ \frac{m_b^2 \tan^2\beta - m_t^2 \cot^2\beta}{m_W^2}  
	\Big)
\Big]
\end{multline}

\begin{multline}
g[\Stop_2,\Stop_2,H^+,H^-] = \frac{ g_2^2}{2} 
\Big[ - (\frac{1}{2} + \frac{1}{6} t_w^2 ) \cos 2\beta 
      - \frac{m_b^2}{m_W^2} \tan^2\beta\\
      + \cos^2\theta_{\Stop} 
	\Big( (\frac{1}{2} + \frac{5}{6} t_w^2) \cos 2\beta 
	+ \frac{m_b^2 \tan^2\beta - m_t^2 \cot^2\beta}{m_W^2}  
	\Big)
\Big]
\end{multline}

\begin{multline}
g[\Sbot_1,\Sbot_1,H^+,H^-] = \frac{ g_2^2}{2} 
\Big[ - \frac{1}{3} t_w^2  \cos 2\beta 
      - \frac{m_b^2}{m_W^2} \tan^2\beta\\
      + \cos^2\theta_{\Sbot} 
	\Big( (\frac{1}{2} + \frac{1}{6} t_w^2) \cos 2\beta 
	+ \frac{m_b^2 \tan^2\beta - m_t^2 \cot^2\beta}{m_W^2}  
	\Big)
\Big]
\end{multline}

\begin{multline}
g[\Sbot_2,\Sbot_2,H^+,H^-] = \frac{ g_2^2}{2} 
\Big[ (\frac{1}{2} - \frac{1}{6} t_w^2)  \cos 2\beta 
      - \frac{m_t^2}{m_W^2} \cot^2\beta\\
      - \cos^2\theta_{\Sbot} 
	\Big( (\frac{1}{2} + \frac{1}{6} t_w^2) \cos 2\beta 
	+ \frac{m_b^2 \tan^2\beta - m_t^2 \cot^2\beta}{m_W^2}  
	\Big)
\Big]
\end{multline}


\section{Feynman graphs}\label{feynman-graphs}

\begin{fmffile}{paper-graphs}

Feynman Graphs with opposite direction of charge flow are not 
depicted but indicated by dots.

\subsection{Squark-Graphs}\label{sq-graphs}

\begin{center}

\begin{tabular}[2]{cc}
\begin{picture} (60,30)(0,-1)
\begin{fmfgraph*}(60,28)
\fmfleft{i1,i2}
\fmfright{o1,o2}
\fmflabel{$g$}{i1}
\fmflabel{$g$}{i2}
\fmflabel{$H^+$}{o1}
\fmflabel{$H^-$}{o2}
\fmf{gluon}{i1,v1}
\fmf{gluon}{i2,v1}
\fmf{scalar,left,tension=.5,label=$\Stop_i/\Sbot_i$,l.side=left}{v1,v2}
\fmf{scalar,left,tension=.5,label=$\Stop_i/\Sbot_i$,l.side=left}{v2,v1}
\fmf{dashes,tension=.8,label=$H^0_1/H^0_2$,l.side=left}{v2,v3}
\fmf{scalar}{v3,o2}
\fmf{scalar}{o1,v3}
\fmfdot{v1,v2,v3}
\end{fmfgraph*}
\end{picture}
&
\begin{picture} (60,30)(-5,-1)
\begin{fmfgraph*}(50,28)
\fmfleft{i1,i2}
\fmfright{o1,o2}
\fmf{gluon}{i1,v1}
\fmf{gluon}{i2,v2}
\fmf{scalar,tension=.3,label=$\Stop_i/\Sbot_i$,l.side=left}{v1,v2}
\fmf{scalar,tension=.3,label=$\Stop_i/\Sbot_i$,l.side=left}{v2,v3}
\fmf{scalar,tension=.3,label=$\Stop_i/\Sbot_i$,l.side=left}{v3,v1}
\fmf{dashes,tension=.6,label=$H^0_1/H^0_2$,l.side=left}{v3,v4}
\fmf{scalar}{v4,o2}
\fmf{scalar}{o1,v4}
\fmfdot{v1,v2,v3,v4}
\end{fmfgraph*}
\end{picture}
\hspace {1cm} + $\cdots$
\end{tabular}

\begin{tabular}[2]{cc}
\begin{picture} (60,30)(-10,-1)
\begin{fmfgraph*}(40,28)
\fmfleft{i1,i2}
\fmfright{o1,o2}
\fmf{gluon}{i1,v1}
\fmf{gluon}{i2,v1}
\fmf{scalar,left,tension=.5,label=$\Stop_i/\Sbot_i$,l.side=left}{v1,v2}
\fmf{scalar,left,tension=.5,label=$\Stop_i/\Sbot_i$,l.side=left}{v2,v1}
\fmf{scalar}{v2,o2}
\fmf{scalar}{o1,v2}
\fmfdot{v1,v2}
\end{fmfgraph*}
\end{picture}
&
\begin{picture} (60,30)(-10,-1)
\begin{fmfgraph*}(40,28)
\fmfleft{i1,i2}
\fmfright{o1,o2}
\fmf{gluon}{i1,v1}
\fmf{gluon}{i2,v2}
\fmf{scalar,tension=.3,label=$\Stop_i/\Sbot_i$,l.side=left}{v1,v2}
\fmf{scalar,tension=.5,label=$\Stop_i/\Sbot_i$,l.side=left}{v2,v3}
\fmf{scalar,tension=.5,label=$\Stop_i/\Sbot_i$,l.side=left}{v3,v1}
\fmf{scalar}{v3,o2}
\fmf{scalar}{o1,v3}
\fmfdot{v1,v2,v3}
\end{fmfgraph*}
\end{picture}
\hspace {1cm} + $\cdots$
\end{tabular}

\begin{tabular}[2]{cc}
\begin{picture} (60,30)(-10,-1)
\begin{fmfgraph*}(40,28)
\fmfleft{i1,i2}
\fmfright{o1,o2}
\fmf{gluon}{i1,v1}
\fmf{gluon}{i2,v1}
\fmf{scalar,tension=.4,label=$\Stop_i$,l.side=right}{v1,v2}
\fmf{scalar,tension=.3,label=$\Sbot_i$,l.side=right}{v2,v3}
\fmf{scalar,tension=.4,label=$\Stop_i$,l.side=right}{v3,v1}
\fmf{scalar}{o1,v2}
\fmf{scalar}{v3,o2}
\fmfdot{v1,v2,v3}
\end{fmfgraph*}
\end{picture}
&
\begin{picture} (60,30)(-10,-1)
\begin{fmfgraph*}(40,28)
\fmfleft{i1,i2}
\fmfright{o1,o2}
\fmf{gluon}{i1,v1}
\fmf{gluon}{i2,v2}
\fmf{scalar,tension=.55,label=$\Stop_j$,l.side=right}{v1,v4}
\fmf{scalar,tension=.55,label=$\Sbot_i$,l.side=right}{v4,v3}
\fmf{scalar,tension=.55,label=$\Stop_j$,l.side=right}{v3,v2}
\fmf{scalar,tension=.55,label=$\Stop_j$,l.side=right}{v2,v1}
\fmf{scalar}{o1,v4}
\fmf{scalar}{v3,o2}
\fmfdot{v1,v2,v3,v4}
\end{fmfgraph*}
\end{picture}
\hspace {1cm} + $\cdots$
\end{tabular}

\begin{tabular}[2]{cc}
\begin{picture} (60,30)(-10,-1)
\begin{fmfgraph*}(40,28)
\fmfleft{i1,i2}
\fmfright{o1,o2}
\fmf{gluon}{i1,v1}
\fmf{gluon}{i2,v1}
\fmf{scalar,tension=.4,label=$\Sbot_i$,l.side=left}{v1,v3}
\fmf{scalar,tension=.3,label=$\Stop_i$,l.side=left}{v3,v2}
\fmf{scalar,tension=.4,label=$\Sbot_i$,l.side=left}{v2,v1}
\fmf{scalar}{o1,v2}
\fmf{scalar}{v3,o2}
\fmfdot{v1,v2,v3}
\end{fmfgraph*}
\end{picture}
&
\begin{picture} (60,30)(-10,-1)
\begin{fmfgraph*}(40,28)
\fmfleft{i1,i2}
\fmfright{o1,o2}
\fmf{gluon}{i1,v1}
\fmf{gluon}{i2,v2}
\fmf{scalar,tension=.55,label=$\Sbot_i$,l.side=left}{v1,v2}
\fmf{scalar,tension=.55,label=$\Sbot_i$,l.side=left}{v2,v3}
\fmf{scalar,tension=.55,label=$\Stop_j$,l.side=left}{v3,v4}
\fmf{scalar,tension=.55,label=$\Sbot_i$,l.side=left}{v4,v1}
\fmf{scalar}{o1,v4}
\fmf{scalar}{v3,o2}
\fmfdot{v1,v2,v3,v4}
\end{fmfgraph*}
\end{picture}
\hspace{1cm} + $\cdots$
\end{tabular}

\begin{tabular}[1]{c}
\begin{picture} (40,30)(0,-1)
\begin{fmfgraph*}(40,28)
\fmfleft{i1,i2}
\fmfright{o1,o2}
\fmf{gluon}{i2,v2}
\fmf{phantom}{i1,v1}
\fmf{scalar,tension=.55,label=$\Sbot_i$,l.side=left}{v1,v4}
\fmf{scalar,tension=.55,label=$\Sbot_i$,l.side=right}{v4,v3}
\fmf{scalar,tension=.55,label=$\Stop_j$,l.side=right}{v3,v2}
\fmf{scalar,tension=.55,label=$\Stop_j$,l.side=right}{v2,v1}
\fmf{scalar}{v3,o2}
\fmf{phantom}{v4,o1}
\fmffreeze
\fmf{gluon}{i1,v4}
\fmf{scalar}{o1,v1}
\fmfdot{v1,v2,v3,v4}
\end{fmfgraph*}
\end{picture}
\vspace{1cm} + $\cdots$
\end{tabular}

\end{center}

\subsection{Quark graphs}\label{q-graphs}

\begin{center}

\begin{tabular}[2]{cc}
\begin{picture} (60,30)(-10,-1)
\begin{fmfgraph*}(40,28)
\fmfleft{i1,i2}
\fmfright{o1,o2}
\fmf{gluon}{i1,v1}
\fmf{gluon}{i2,v2}
\fmf{fermion,tension=.55,label=b,l.side=left}{v1,v2}
\fmf{fermion,tension=.55,label=b,l.side=left}{v2,v3}
\fmf{fermion,tension=.55,label=t,l.side=left}{v3,v4}
\fmf{fermion,tension=.55,label=b,l.side=left}{v4,v1}
\fmf{scalar}{o1,v4}
\fmf{scalar}{v3,o2}
\fmfdot{v1,v2,v3,v4}
\end{fmfgraph*}
\end{picture}
&
\begin{picture} (60,30)(-10,-1)
\begin{fmfgraph*}(40,28)
\fmfleft{i1,i2}
\fmfright{o1,o2}
\fmf{gluon}{i1,v1}
\fmf{gluon}{i2,v2}
\fmf{fermion,tension=.55,label=t,l.side=left}{v1,v2}
\fmf{fermion,tension=.55,label=t,l.side=left}{v2,v3}
\fmf{fermion,tension=.55,label=b,l.side=right}{v3,v4}
\fmf{fermion,tension=.55,label=t,l.side=left}{v4,v1}
\fmf{phantom}{o1,v4}
\fmf{phantom}{v3,o2}
\fmffreeze
\fmf{scalar}{v4,o2}
\fmf{scalar}{v3,o1}
\fmfdot{v1,v2,v3,v4}
\end{fmfgraph*}
\end{picture}
\end{tabular}

\begin{tabular}[2]{cc}
\begin{picture} (60,30)(-10,-1)
\begin{fmfgraph*}(40,28)
\fmfleft{i1,i2}
\fmfright{o1,o2}
\fmf{gluon}{i2,v2}
\fmf{phantom}{i1,v1}
\fmf{fermion,tension=.55,label=b,l.side=left}{v1,v2}
\fmf{fermion,tension=.55,label=b,l.side=left}{v2,v3}
\fmf{fermion,tension=.55,label=t,l.side=left}{v3,v4}
\fmf{fermion,tension=.55,label=t,l.side=right}{v4,v1}
\fmf{scalar}{v3,o2}
\fmf{phantom}{o1,v4}
\fmffreeze
\fmf{gluon}{i1,v4}
\fmf{scalar}{v1,o1}
\fmfdot{v1,v2,v3,v4}
\end{fmfgraph*}
\end{picture}
&
\begin{picture} (60,30)(-5,-1)
\begin{fmfgraph*}(50,28)
\fmfleft{i1,i2}
\fmfright{o1,o2}
\fmf{gluon}{i1,v1}
\fmf{gluon}{i2,v2}
\fmf{fermion,tension=.3,label=t/b,l.side=left}{v1,v2}
\fmf{fermion,tension=.3,label=t/b,l.side=left}{v2,v3}
\fmf{fermion,tension=.3,label=t/b,l.side=left}{v3,v1}
\fmf{dashes,tension=.6,label=$H^0_1/H^0_2$,l.side=left}{v3,v4}
\fmf{scalar}{v4,o2}
\fmf{scalar}{o1,v4}
\fmfdot{v1,v2,v3,v4}
\end{fmfgraph*}
\end{picture}
\end{tabular}

\vspace{1cm} + $\cdots$

\end{center}

\end{fmffile}


\newpage

\section*{Figures}

\newpage


\begin{thebibliography}{ABCDEF}

\def\ap#1#2#3{{\it Ann. Phys. }{\bf #1~}(19#2)~#3}
\def\app#1#2#3{{\it Acta Phys. Polon. }{\bf #1~}(19#2)~#3}
\def\arnps#1#2#3{{\it Annu. Rev. Nucl. Part. Sci. }{\bf #1~}(19#2)~#3}
\def\fp#1#2#3{{\it Fortschr. Phys. }{\bf #1~}(19#2)~#3}
\def\hepph#1{{\bf hep-ph}/#1}
\def\ijmp#1#2#3{{\it Int. Jour. Mod. Phys. }{\bf #1~}(19#2)~#3}
\def\jetp#1#2#3{{\it JETP Lett. }{\bf #1~}(19#2)~#3}
\def\mpl#1#2#3{{\it Mod. Phys. Lett. }{\bf #1~}(19#2)~#3}
\def\npb#1#2#3{{\it Nucl. Phys. }{\bf B#1~}(19#2)~#3}
\def\plb#1#2#3{{\it Phys. Lett. }{\bf B#1~}(19#2)~#3}
\def\prd#1#2#3{{\it Phys. Rev. }{\bf D#1~}(19#2)~#3}
\def\prep#1#2#3{{\it Phys. Rep. }{\bf #1~}(19#2)~#3}
\def\prl#1#2#3{{\it Phys. Rev. Lett. }{\bf #1~}(19#2)~#3}
\def\ptp#1#2#3{{\it Prog. Theor. Phys. }{\bf #1~}(19#2)~#3}
\def\rmp#1#2#3{{\it Rev. Mod. Phys. }{\bf #1~}(19#2)~#3}
\newcommand{\nc}[3]{{\it Nuovo Cim. }{\bf #1} (19#2) #3}
\newcommand{\ncl}[3]{{\it Nuovo Cim. Lett. }{\bf #1} (19#2) #3}
\def\rnc#1#2#3{{\it Riv. Nuovo Cim. }{\bf #1~}(19#2)~#3}
\def\sjnp#1#2#3{{\it Sov. J. Nucl. Phys. }{\bf #1~}(19#2)~#3}
\def\zpc#1#2#3{{\it Z. Phys. }{\bf C#1~}(19#2)~#3}
\def\zp#1#2#3{{\it Z. Phys. }{\bf #1~}(19#2)~#3}
\newcommand{\prold}[3]{{\sl Phys. Rev.} {\bf #1} (19#2) #3}
\newcommand{\pl}[3]{{\sl Phys. Lett.} {\bf #1} (19#2) #3}


\bibitem{EHLQ} E.~Eichten, I.~Hinchliffe, K.~Lane, C.~Quigg, 
        \rmp{56}{84}{579}; \\
      N.G.~Deshpande, X.~Tata, D.A.~Dicus, 
        \prd{29}{84}{1527}. 



\bibitem{willenbrock} S.S.D.~Willenbrock, 
        \prd{35}{87}{173}.


\bibitem{KPSZ} A.~Krause, T.~Plehn, M.~Spira, P.M.~Zerwas, 
        \npb{519}{98}{85}.

\bibitem{kniehl}
        A.A. Barrientos Bendez\'u, B.A. Kniehl, hep-ph/9908385

\bibitem{ERZ} J.~Ellis, G.~Ridolfi, F.~Zwirner, 
        \plb{262}{91}{477}; \\
        A.~Dabelstein, 
        \npb{456}{95}{25}.


\bibitem{Diaz} M.A. Diaz, H.E. Haber,
	\prd{45}{92}{4246}


\bibitem{HHG} J.F.~Gunion, H.E.~Haber, G.~Kane, S.~Dawson, 
        {\em The Higgs Hunter's Guide }, Addison-Wesley Publishing Company, 
        1990.


\bibitem{GB} E.W.N.~Glover, J.J.~van der Bij, 
        \npb{309}{88}{282}.


\bibitem{PSZ} T.~Plehn, M.~Spira, P.M.~Zerwas, 
        \npb{479}{96}{46}



\bibitem{pQCD-lect} 
        G.~Sterman et al, 
        \rmp{67}{95}{1}. 


\bibitem{MRSG} A.D.~Martin, W.J.~Stirling, R.G.~Roberts, 
        \plb{354}{95}{155}.


\end{thebibliography}
\end{document}